\documentclass{llncs}
\usepackage{graphicx}
\usepackage[table,xcdraw]{xcolor}
\usepackage{caption}
\usepackage{subcaption}
\usepackage{hyperref}
\usepackage{multirow}
\usepackage[table,xcdraw]{xcolor}
\begin{document}

\title{On Time and Space: An Experimental Study on Graph Structural and Temporal Encodings} 

%
\titlerunning{Experimental Study on Graph Structural and Temporal Encodings}
%
\author{Velitchko Filipov\inst{1}\orcidID{0000-0001-9592-2179} \and
Alessio Arleo\inst{1}\orcidID{0000-0003-2008-3651} \and
Markus B\"ogl\inst{1}\orcidID{0000-0002-8337-4774} \and Silvia Miksch\inst{1}\orcidID{0000-0003-4427-5703}}
%
\authorrunning{V. Filipov et al.}
%
\institute{TU Wien, Vienna, Austria\\
\email{\{firstname.lastname\}@tuwien.ac.at}}
%
\maketitle              

\begin{abstract}
Dynamic networks reflect temporal changes occurring to the graph's structure and are used to model a wide variety of problems in many application fields. We investigate the design space of dynamic graph visualization along two major dimensions: the network structural and temporal representation. Significant research has been conducted evaluating the benefits and drawbacks of different structural representations for static graphs, however, few extend this comparison to a dynamic network setting.
We conduct a study where we assess the participants' response times, accuracy, and preferences for different combinations of the graph's structural and temporal representations on typical dynamic network exploration tasks, with and without support of common interaction methods.
Our results suggest that matrices provide better support for tasks on lower-level entities and basic interactions require longer response times while increasing accuracy. Node-link with auto animation proved to be the quickest and most accurate combination overall, while animation with playback control the most preferred temporal encoding.  
\keywords{User Study \and Evaluation \and Time-oriented Data \and Graphs \& Networks.}
\end{abstract}

\section{Introduction}
\label{sec:introduction}
The increased availability of time-dependent datasets contributed to the rise of research interest toward dynamic network visualization, nowadays considered a mature and thriving research field~\cite{beck2017}. Kerracher et al.~\cite{kerracher2014} define a two-dimensional design space for dynamic network visualization: \textit{structural representation} (how the graph's topology is represented) and \textit{temporal encoding} (how time and, consequently, the graph temporal dynamics are illustrated). 
This two-dimensional design space is expressive enough to characterize the majority of existing dynamic network visualization approaches.

There is extensive literature on studies designed to evaluate different graph representations for typical exploration tasks on static networks. Similar studies have been conducted for dynamic approaches, however mostly focused on node-link diagrams coupled with different temporal encodings (see Section~\ref{sec:related_work}). This also comes as a consequence of the limited number of dynamic network visualization approaches that have matrices as their base graph representation~\cite{beck2017} (see, e.g., ~\cite{henry2007matlink,rufiange2014animatrix}).
Therefore, empirical evidence about the performance and preference for different dynamic network visualization approaches in our design space
is still scattered between different studies, experimental settings, procedures, and tasks.
Similarly, existing user studies in this context incorporate simple interaction methods to support the network exploration (see, e.g.,~\cite{archambault2013map,okoe2018}), however, their effect on the participants' experience has not been fully investigated, thus motivating the need for broader and rigorous experimentation. 

\textbf{Our Contribution.} In this paper we design, conduct, and discuss the results of an experimental study aimed at assessing and comparing different dynamic visualization approaches centered around combinations of graph representation (node-link and adjacency matrix), temporal encoding (juxtaposition, superimposition, animation with playback control, and auto animation), and interaction support for offline dynamic graph visualization. We conduct a statistical analysis of the study results and condense our findings in a concise discussion meant to support further studies and design of dynamic network visualization techniques. 

\section{Related Work}
\label{sec:related_work}

We outline recent related studies conducted along the two dimensions of the design space introduced by Kerracher et al.~\cite{kerracher2014}, focusing on user studies.

\noindent\textbf{Structural Representations.} In graph drawing literature, several studies assess the readability, task performance, and effects of aesthetic criteria on human cognition of different graph structural encodings (e.g.,~\cite{bennett2007,di2021user,donghao2019,ghoniem2004,okoe2018,okoe2019,purchase1998,purchase2002}). 
Ghoniem et al.~\cite{ghoniem2004} evaluate, in a controlled experiment, the readability of graphs when represented as node-link diagrams compared to adjacency matrices on generic graph tasks. Their findings suggest that the ability of either visualization to support typical exploration tasks depends on the size and density of the network; the authors concluded how matrix-based techniques were under-exploited, despite their proved potential with larger and denser networks. 
Okoe et al.~\cite{okoe2018,okoe2019} conduct further comparative evaluations between node-link and matrix representations on a large scale ($\sim800$ participants). 
Their results show that node-link diagrams better support memorability and connectivity tasks. Matrices have quicker and more accurate results for tasks that involve finding common neighbors and group tasks (i.e., involving clusters).
Concurrently, Ren et al.~\cite{donghao2019} conduct a large scale study ($\sim600$ participants) comparing the readability of node-link diagrams against two different sorting variants of matrix representations on small to medium social networks ($\sim50$ nodes). Their findings do not differ significantly from the ones by Okoe et al.~\cite{okoe2018}, suggesting that node-link provided a better implicit understanding of the network, with lower response times and higher accuracy than matrices. However, the gap between the two tended to reduce as the size of the graph increased. 

\noindent\textbf{Temporal Encodings.} One of the most studied problems concerning dynamic network visualization, is the ability of participants to retain a ``mental map'' of the graph while investigating its evolution~\cite{archambault2011,archambault2012,archambault2013map,archambault16,purchase2006important}. Archambault and Purchase investigate the effect of drawing stability on the node-link graph representation coupled with animation and small multiples~\cite{archambault2013map,archambault16}. Drawing stability proved to have a positive effect on task performance, with animation able to improve over timeline in low stability scenarios.
Ghani et al.~\cite{ghani2012} investigate the perception of different visual graph metrics on animated node-link diagrams.
Results suggest that animation speed and target separation have the most impact on performance for event sequencing tasks. Linhares et al.~\cite{linhares2021comparative} compare four different approaches for visualization of dynamic networks, namely the \texttt{Massive Sequence View}~\cite{van2013dynamic} (timeline-based), the \texttt{Temporal Activity Map}~\cite{linhares2017dynetvis}, and animated node-link and matrix diagrams. While all techniques reached satisfactory results, the animated node-link was the favorite choice of the participants. Even though matrix-based approaches are included in this study, it does not exhaustively cover all the possible combinations of our design space.
Filipov et al.~\cite{filipov2021exploratory} conduct an exploratory study comparing different combinations of structural and temporal representations.
The results suggest that tasks with matrices were completed quicker and more accurately, the participants preferred matrices with superimposition, and juxtaposition was among the least preferred approaches.
However, these results require further formal investigation.
Overall, related literature shows that the perception of different temporal encodings has been mainly investigated on node-link diagrams, with few papers focusing on the other combinations of structural and temporal encodings.
In this sense, our paper constitutes an effort in understanding whether the differences between node-link and matrix representations still hold in a dynamic scenario, what is the efficacy of the temporal representations, and how effective (and how important) is it to include interactions when designing such approaches.

\section{Dynamic Graph Visualization Design Space}
\label{sec:design_space}
We refer to a dynamic graph $\Gamma$ as a sequence of individual graphs each one representing its state at a specific point in time: $\Gamma = (G_1, G_2,..., G_k)$;
we denote the individual $G_x$ as a dynamic graph \textit{timeslice}. 
We now briefly describe the different structural 
and temporal encodings, along with the interactions included in the scope of our experiment, detailing their implementation. 

\subsection{Network Structural Encoding}
The structural dimension focuses on the challenges of laying out a graph to visually present the relationships between elements in an understandable, accurate, and usable manner~\cite{kerracher2014}. 
\textbf{Node-Link (NL)} diagrams 
present the relational structure of the graph using lines to connect the entities that are depicted using circles, whose coordinates on the plane are computed using specialized algorithms. 
In our study, we compute the NL layouts using the force-directed implementation of \textit{d3js}~\cite{bostock2011} considering all the timeslices simultaneously instead of on a per-frame basis. This process of \textit{aggregation}~\cite{brandes2011quantitative} is simple to implement and provides a stable layout throughout the sequence of timeslices, at the expense of the quality of individual layouts. We refer to the following for a broader discussion on dynamic network layout algorithms~\cite{arleo2022event,baur2001visone,beck2017,brandes2003visual,collberg2003system,crnovrsanin2017incremental,erten2003graphael,erten2004exploring,simonetto2018event}.
\textbf{Adjacency Matrices (M)} visualize the network as an $n \times n$ table. 
A non-zero value in the cell indicates the presence of an edge between the nodes identified by the corresponding row and column. 
In our study we order the rows and columns alphabetically according to the node's label. More advanced reordering methods exist~\cite{behrisch2016matrix}, however, matrix reordering is still under-investigated in a dynamic context and we decided to exclude this aspect from the study design.

\begin{figure}[t]
    \centering
    \includegraphics[width=\textwidth]{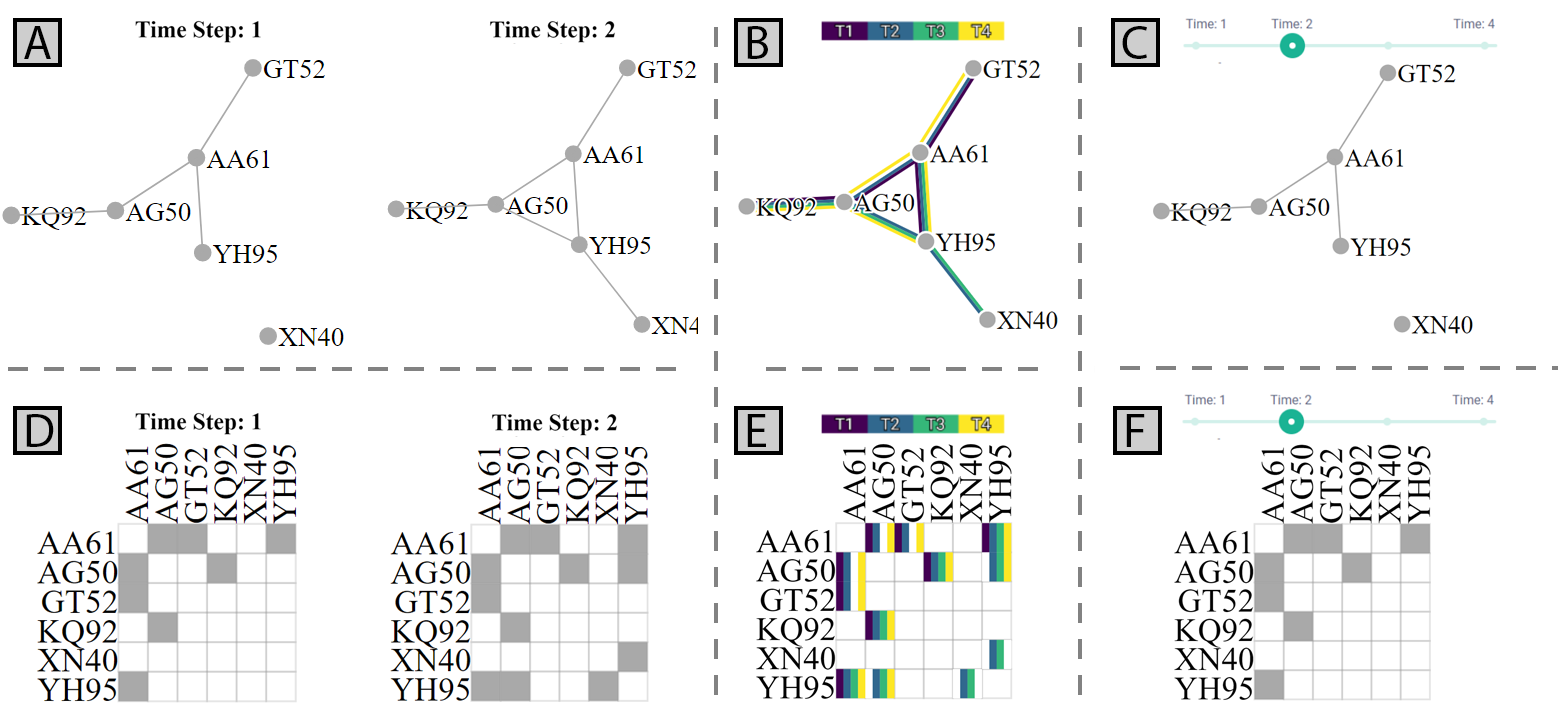}
    \caption{Network structural and temporal encodings: Juxtaposition (A,D), Superimposition(B,E), and Animation with Playback Controls (C,F)}
    \label{fig:temporal_encodings}
\end{figure}

\subsection{Network Temporal Encoding}
In dynamic networks the temporal dimension plays an important role in the analysis process and requires special attention to enable effective exploration and better understand the behavior of the network~\cite{miksch2014}. \textbf{Superimposition (SI)}
encodes the temporal dimension of the network in the same screen space by overlaying the timeslices (see, e.g.,~\cite{brandes2003visual,erten2004exploring}) or making use of explicit encoding (see., e.g., ~\cite{gleicher2011,javed2012}). 
In our study we represent the temporal information in SI using colorblind-friendly color palettes~\cite{viridis}. In NL, we generate multiple parallel edges between the nodes, one for each timeslice where the edge is present, and color-code them individually. In M we subdivide each cell uniformly into rectangles, each representing the existence of that edge during that timeslice, which are colored similarly (see Figure~\ref{fig:temporal_encodings} B-E). 
\textbf{Juxtaposition (JP)} represents the graph's temporal dynamics as distinct layouts, each with dedicated screen space, similar to
the small multiples approach by Tufte~\cite{tufte83} (see Figure~\ref{fig:temporal_encodings}A,D). In our study we generate one diagram per timeslice and arrange them adjacent to each other.
\textbf{Animation with Playback Control (ANC)}
uses a time slider to control the state of the animation and move to any of the available timeslices in no particular order (see Figure~\ref{fig:temporal_encodings}C,F).
This enables for a more fine and controlled exploration and analysis compared to animation, where speed and time progression is typically fixed.
\noindent \textbf{Auto Animation (AN)}
depicts the change of the graph over time as smooth transitions between subsequent timeslices.
Differently from ANC, with AN it is not possible to skip forward or navigate backward in time and it automatically goes over each of the timeslices in a sequence.

\subsection{Interactions}\label{se:interactions}

The interactions we implement are meant to support the network exploration.
The following apply regardless of the temporal encoding: (i) zooming and panning (both for M and NL); (ii) 
hovering over a M cell highlights its corresponding row and column; (iii) in NL, nodes can be moved by dragging in order to de-clutter some denser areas of the drawing. Moreover, for AN only and regardless of the structural representation, the time between consecutive timeslices can be increased (7 sec maximum) or decreased (1 sec minimum).
This selection should not favor any specific combination of structural and temporal encoding techniques over the others.
Zooming, panning, and node rearrangement are commonly available in graph exploration software, like \textit{Gephi}~\cite{bastian2009gephi}. M mouse-over was also used by Okoe et al.~\cite{okoe2018}. AN speed could also be manipulated in the study by Archambault and Purchase~\cite{archambault2012}.

\section{Study Design}
\label{sec:study_design}

In this section we present the study design, including our tasks, research  hypotheses, stimuli, and study procedure.

\subsection{Tasks and Research Hypotheses}\label{sec:eval_tasks} 
\begin{table}[!hb]
\centering
\begin{tabular}{|
>{\columncolor[HTML]{B7B7B7}}p{0.5cm}|p{5.75cm}|p{5.75cm}|}
\hline
\textbf{T}  & \cellcolor[HTML]{B7B7B7}\textbf{Low-level}                                                              & \cellcolor[HTML]{B7B7B7}\textbf{High-level}                                                                                        \\ \hline \hline
\textbf{T1} & At which time step is the relationship between \texttt{\{source\}} and \texttt{\{target\}} introduced for the first time? & At which time step does the clique between \texttt{\{nodes\}} appear for the first time?                                                        \\ \hline \hline
\textbf{T2} & Sum up the  changes (additions and removals) of \texttt{\{node\}}’s degree across all time steps.              & Calculate the change of the clique’s size between \texttt{\{nodes\}} across all time steps.                                                     \\ \hline \hline
\textbf{T3} & At which time step does the node \texttt{\{node\}} have its highest degree?                                      & Consider the set of nodes \texttt{\{nodes\}}. Find the size of the largest maximal clique across all the time steps between the given nodes. \\ \hline
\end{tabular}
    \caption{The test questions (trials), per task (rows) and entity type (columns).}
    \label{tab:tasks}
\end{table}
\noindent\textbf{Tasks.} The tasks used in our experiment are available in Table~\ref{tab:tasks}.

\noindent\textbf{Rationale.} We picked one task for each category of temporal feature in the taxonomy proposed by Ahn et al.~\cite{ahn2014}, namely, \textit{Individual Temporal Features} (\textbf{T1}), \textit{Rate of changes} (\textbf{T2}), and \textit{Shape of changes} (\textbf{T3}). 
We selected the most common tasks referenced in the taxonomy and included in our experiment these tasks for both low- (nodes and links) and higher-level (cliques) entities.

\label{sec:hypotheses}

\begin{table}[!h]
\centering
\begin{tabular}{|
>{\columncolor[HTML]{B7B7B7}}p{0.5cm} |p{11.5cm}|}
\hline
\multicolumn{1}{|l|}{\cellcolor[HTML]{B7B7B7}\textbf{H}} & \cellcolor[HTML]{B7B7B7}\textbf{Research Hypothesis}                                                                                                                                                                                 \\ \hline \hline
\textbf{H1}                                              & Matrices have lower response times and higher accuracy for all tasks compared to node-link diagrams, regardless of the temporal encoding.                                                                                            \\ \hline \hline
\textbf{H2}                                              & From all temporal encoding techniques, superimposition has the lowest response times and highest accuracy, regardless of the structural representation.                                                                              \\ \hline \hline
\textbf{H3}                                              & Providing interaction techniques increases the response times but not the accuracy.                                                                                                                                                  \\ \hline \hline
\textbf{H4}                                              & Matrices have lower response times and higher accuracy for tasks on low-level entities and node-link diagrams have lower response times and higher accuracy for tasks on higher-level entities, regardless of the temporal encoding. \\ \hline \hline
\textbf{H5}                                              & The combination of matrices with superimposition results in the lowest response times and highest accuracy compared to other combinations of network structural and temporal encoding.                                               \\ \hline
    \end{tabular}
    
    \caption{The research hypotheses that were evaluated in our experiments.}
    \label{tab:hypotheses}
\end{table}
\noindent\textbf{Hypotheses.} We base our research hypotheses on the proposed tasks and we report them in Table~\ref{tab:hypotheses}.

\noindent\textbf{Rationale.} Hypotheses \textbf{H1}, \textbf{H2}, and \textbf{H5} are derived from the results of a previous exploratory study~\cite{filipov2021exploratory} (see also Section~\ref{sec:related_work}).
While the focus of this experiment is centered around the \textit{visual} encoding combinations within our design space, \textbf{H3} is intended to investigate the effects of common interactions techniques in this context. We argue that they might increase the response times over visual inspection alone, 
but without significant impact on accuracy.
In \textbf{H4} 
we conjecture that following the evolution of a cluster or clique is more difficult with M compared to NL, as the participant must track several elements at once. We assume this would be easier to achieve with NL as the nodes are drawn closer together.

\subsection{Experiment Setting}\label{sec:eval_setting}

\noindent\textbf{Stimuli.} We generated 24 different scale-free random~\cite{bollobas2003directed}  graphs ($35 \leq |V| \leq 45, 46 \leq |E| \leq 71 $) with the \textit{NetworkX} python library~\cite{hagberg2008exploring,hagberg2020networkx}. We chose this category of networks as they resemble real-world data examples of scientific interest (e.g., the world-wide-web, authors' co-citation networks~\cite{albert2002statistical}). 
We augmented each graph with 4 timeslices by randomly deleting edges from the original input graph to simulate temporal dynamics (at each subsequent timeslice the edges were added back and a new set was selected for removal). 
Finally, we split the datasets into two different types: 12 graphs with cliques and 12 without. Cliques were artificially introduced in the graphs by choosing 5 random nodes which were fully connected in one or more of the graph timeslices.
The size of the graphs is comparable with the majority of empirical studies on graph visualization~\cite{donghao2019,yoghourdjian2018}.

\noindent\textbf{Trials.} Each of the tasks is applied to all combinations of structural and temporal encodings of interest in our study (see Section~\ref{sec:design_space})
resulting in 48 unique trials: $3 (task\ types) \times 2 (entity\ types) \times 2 (network\  encodings) \times 4 (temporal\ encodings)$. The order of the trials during the study is randomized in order to mitigate learning effects. The participants take part in the online experiment by completing the trials prepared using SurveyJS~\cite{surveyjs}.

\noindent\textbf{Study Design.} Our experiment follows a between-subject arrangement: all participants complete the same entire set of 48 trials on the same graphs, but are exposed to one of two conditions, either \textit{without} (Group \textit{A}) or \textit{with} (Group \textit{B}) the support of the interactions discussed in Section~\ref{se:interactions}. Participants are assigned to one of the two groups when they first access the online experiment, with a 75\% probability of being assigned to Group \textit{B}. As only one hypothesis (\textbf{H3}) deals with the group subdivision, we design the experiment to have a higher number of participants with interaction support.  
For each trial we ask the participant to provide
a confidence score of their answer using a 5 point Likert scale (1 least confident - 5 most confident). 
At the end of the experiment, the participants express their thoughts in text about the encoding combinations they encountered and rank them
on a 5 point Likert scale (1 least preferred - 5 most preferred).%

\noindent\textbf{Participants}. For our study, we enrolled students part of a graduate course on information visualization design. 
To ensure that participants had a sufficient level of knowledge on the topic, we gave an introductory lecture about the visualizations and the experiment modalities. Participation was optional and its performance did not impact the final grade of the students. 
The online setting was necessary to guarantee a sufficient number of participants, while ensuring a safe social-distancing protocol. However, this also meant giving up control on the experiment environment.

\section{Study Results}
\label{sec:results}
We received a total of 76 submissions from as many participants, of which we removed 8 that were trying to game the experiment. This resulted in a final set of 68 valid submissions that were used as the basis of our analysis. Further information and the complete set of results can be found in the Appendix.

\subsection{Analysis Approach}\label{se:stat_analysis}
For each question of our study, we collected the participants' answers, their corresponding response times, and confidence values. 
We ignored the group subdivision (Group \textit{A} and \textit{B}) for hypotheses which did not focus on the presence of interactions (all except \textbf{H3}, see  Section~\ref{sec:hypotheses}), as ANOVA tables do not show a statistically significant interaction effect between the independent variables for \textbf{H1}, \textbf{H2}, \textbf{H4}, \textbf{H5} (for more information we refer to the Appendix). 

We conduct our analysis as follows, supported by Python libraries for statistical analysis~\cite{harris2020array,pingouin,2020SciPy-NMeth}. We consider the structural and temporal encoding, the task type, entity type, and the groups (Group \textit{A} and \textit{B}) as \textit{independent variables}, the response times and accuracy are taken as \textit{dependent variables}. 
As the group subdivision is not even (25-75), we choose methods that are robust against these unbalanced designs \cite{backhaus2015fortgeschrittene,bortz2013statistik,hedderich2016angewandte,weiss2005basiswissen}. For each of the hypotheses, we grouped the data according to the hypothesis and visually inspected response times and accuracy (number of correct answers $\div$ total number of answers).
To remove outliers from the data before the analysis we employed the inter-quantile range (IQR)~\cite{rousseeuw1993alternatives}.
We set the IQR lower ($q_1- 1.5 \cdot \textrm{IQR}$) and upper ($q_2+1.5 \cdot \textrm{IQR}$) bounds at $q_1=0.25$ and $q_2=0.75$ as the outlier cut-off boundaries. This resulted in 116 trials (or 3.43\%) being detected as outliers and omitted from the analysis.

\begin{table}[!ht]
\resizebox{\textwidth}{!}{
\begin{tabular}{|
>{\columncolor[HTML]{B7B7B7}}c |
>{\columncolor[HTML]{D9D9D9}}r |r|r|r|}
\hline
\textbf{Hypothesis}                                   & \multicolumn{1}{c|}{\cellcolor[HTML]{B7B7B7}\textbf{Groups}} & \multicolumn{1}{c|}{\cellcolor[HTML]{B7B7B7}\textbf{MWU}} & \multicolumn{1}{c|}{\cellcolor[HTML]{B7B7B7}\textbf{T-Test}} & \multicolumn{1}{c|}{\cellcolor[HTML]{B7B7B7}\textbf{Binomial}} \\ \hline \hline
\cellcolor[HTML]{B7B7B7}                              & \textit{(NL T1) vs (M T1)}                                   & \textbf{0.0104*$^b$}                                          & \textbf{\textless 0.001***$^b$}                                  & \textbf{0.0013*$^b$}                                               \\ \cline{2-5} 
\cellcolor[HTML]{B7B7B7}                              & \textit{(NL  T2) vs (M T2)}                                  & 0.1579                                                    & \textbf{\textless 0.001***$^b$}                                  & 0.9313                                                         \\ \cline{2-5} 
\multirow{-3}{*}{\cellcolor[HTML]{B7B7B7}\textbf{H1}} & \textit{(NL T3) vs (M T3)}                                   & \textbf{\textless 0.001***$^b$}                               & \textbf{\textless 0.001***$^b$}                                  &        \textbf{0.0022**$^b$}                                               \\ \hline \hline
\cellcolor[HTML]{B7B7B7}                              & \textit{(SI) vs (JP)}                                        & \textbf{\textless 0.001***$^b$}                               & 0.1065                                                       & 0.166                                                          \\ \cline{2-5} 
\cellcolor[HTML]{B7B7B7}                              & \textit{(SI) vs (ANC)}                                       & 0.8662                                                    & 0.1429                                                       & 0.0883                                                         \\ \cline{2-5} 
\multirow{-3}{*}{\cellcolor[HTML]{B7B7B7}\textbf{H2}} & \textit{(SI) vs (AN)}                                        & 0.2766                                                    & 0.7751                                                       & \textbf{\textless 0.001***$^b$}                                    \\ \hline \hline
\textbf{H3}                                           & \textit{(Grp A) vs (Grp B)}                                  & \textbf{\textless 0.001***}                               & \textbf{\textless 0.001***}                                  & \textbf{\textless 0.001***}                                    \\ \hline \hline
\cellcolor[HTML]{B7B7B7}                              & \textit{(M Low) vs (NL Low)}                                 & \textbf{\textless 0.001***$^b$}                               & 0.1392                                                       & \textbf{\textless 0.001***$^b$}                                    \\ \cline{2-5} 
\multirow{-2}{*}{\cellcolor[HTML]{B7B7B7}\textbf{H4}} & \textit{(M High) vs (NL High)}                               & \textbf{\textless 0.001***$^b$}                               & \textbf{\textless 0.001***$^b$}                                  & 0.4321                                                         \\ \hline \hline
\cellcolor[HTML]{B7B7B7}                              & \textit{(M+SI) vs (M+JP)}                                    & \textbf{0.0056**}                                                    & 0.2567                                                       & 0.2424                                                         \\ \cline{2-5} 
\cellcolor[HTML]{B7B7B7}                              & \textit{(M+SI) vs (M+ANC)}                                   & 0.6301                                                    & 0.2989                                                       & 0.0261                                                         \\ \cline{2-5} 
\cellcolor[HTML]{B7B7B7}                              & \textit{(M+SI) vs (M+AN)}                                    & 0.2766                                                    & 0.6328                                                       & 0.0646                                                         \\ \cline{2-5} 
\cellcolor[HTML]{B7B7B7}                              & \textit{(M+SI) vs (NL+SI)}                                   & \textbf{0.0038**$^b$}                                                    & \textbf{\textless 0.001***$^b$}                                                       & 0.449                                                          \\ \cline{2-5} 
\cellcolor[HTML]{B7B7B7}                              & \textit{(M+SI) vs (NL+JP)}                                   & \textbf{\textless 0.001***$^b$}                               & \textbf{\textless 0.001***$^b$}                                  & 0.1389                                                         \\ \cline{2-5} 
\cellcolor[HTML]{B7B7B7}                              & \textit{(M+SI) vs (NL+ANC)}                                  & 0.0088                                                    & \textbf{\textless 0.001***$^b$}                                  & \textbf{\textless 0.001***$^b$}                                    \\ \cline{2-5} 
\multirow{-7}{*}{\cellcolor[HTML]{B7B7B7}\textbf{H5}} & \textit{(M+SI) vs (NL+AN)}                                   & 0.0331                                                    & \textbf{\textless 0.001***$^b$}                                  & \textbf{\textless 0.001***$^b$}                                    \\ \hline
\end{tabular}
}
\caption{The results of the statistical test (p-values) for each hypothesis. We mark the cells with * if $p < 0.05$, ** if $p < 0.01$, *** if $p < 0.001$. If multiple comparisons are performed, $^b$ indicates the Bonferroni correction~\cite{bonferroni1936teoria}.}
\label{tab:results}
\end{table} 

The task response times in our experiment are not normally distributed.
To mitigate this, we perform a Box-Cox transformation~\cite{box1964analysis}. Visual inspection of the quantile-quantile (Q-Q) plots confirmed a normal distribution of the transformed data.
This allows us to run parametric tests, specifically, ANOVA (see Appendix for further information about the ANOVA tables) and T-tests \cite{backhaus2015fortgeschrittene,bortz2013statistik,hedderich2016angewandte,weiss2005basiswissen}. 
The standard ANOVA and T-tests are robust against such skewed distributions~\cite{boneau1960effects,posten1984robustness,schminder2010really}, therefore, 
we rely on them for our analysis as they both have more statistical power than non-parametric tests and detect significant effects if they truly exist. In presence of statistically significant difference ($p\textrm{-value}<0.05$),
we check, with T- and Mann-Whitney-U (MWU) tests, whether the significance held and visually explored the corresponding box plots to come to a conclusion.
To evaluate our hypotheses on accuracy, we also perform Binomial tests to detect statistical significance between the distributions.

\subsection{Quantitative Results}

\begin{figure}[h]
     \centering
     \includegraphics[width=\textwidth]{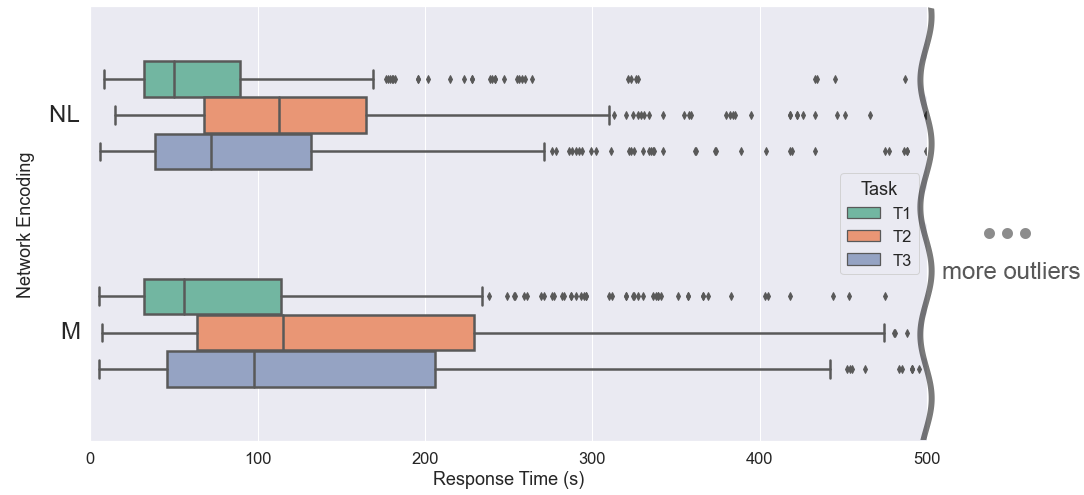}
     \caption{\textbf{H1:} Box plot of response times for NL and M per task.}
     \label{fig:h1_response}
\end{figure}

\noindent\textbf{H1.} We presume based on previous work~\cite{filipov2021exploratory} that M would perform better overall compared to NL for all tasks. Figure~\ref{fig:h1_response} depicts differences in response times between M and NL diagrams per task type. The results (see Table~\ref{tab:results}) indicate that NL is generally faster and more accurate than M. However, when looking at their differences per task we discover for \textbf{T1} that NL is significantly faster than M (NL: 73.49s, M: 97.93s), whereas M proves to be more accurate (NL: 74.9\%, M: 80.7\%).  For \textbf{T2} the T-Test detects a significant difference in response times between NL and M (NL: 133.41s, M: 194.20s), however, in terms of accuracy they both perform similarly (NL: 52.5\%, M: 52.7\%). For \textbf{T3} NL representations significantly outperform M in terms of response times (NL: 107.32s, M: 175.92s) as well as accuracy (NL: 65.7\%, M: 59.4\%). Summarizing, the results suggest
NL
to generally have
the lowest response times and higher accuracy compared to M for the proposed tasks. Thus, our results do not support H1.

\begin{figure}[!htp]
     \centering
     \includegraphics[width=\textwidth]{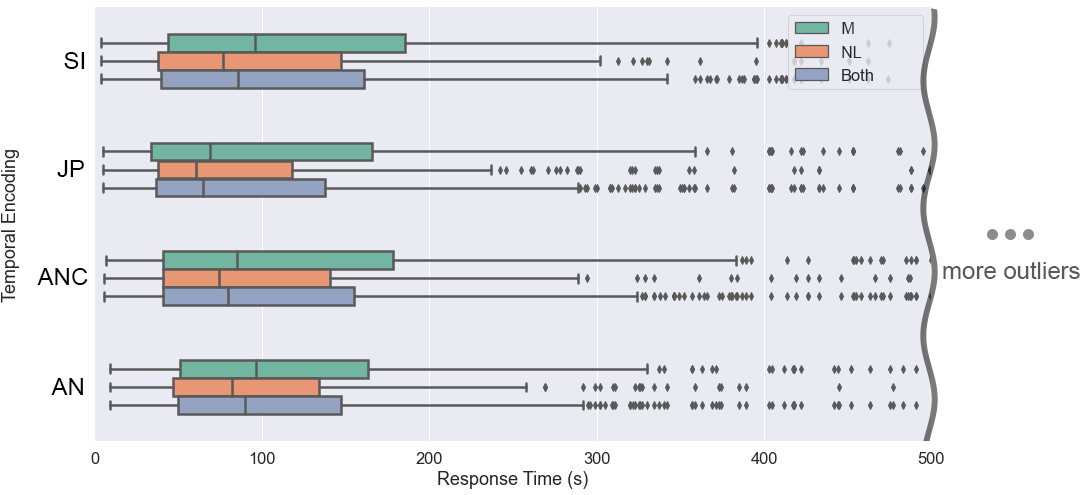}
     \caption{\textbf{H2:} Box plot of response times for temporal and network representations.}
     \label{fig:h2_response}
\end{figure}

\noindent\textbf{H2.} We assume SI to have the lowest response times and highest accuracy out of all the temporal encoding techniques. In our analysis, however, we do not detect any statistical significance in the comparisons shown in Table~\ref{tab:results},
with the only exception being JP, which has considerably lower response times than SI (see Figure~\ref{fig:h2_response}). Concerning response times, JP has the lowest (118.32s), followed by AN (127.76s), SI (129.69s), and ANC (141.35s). We also run a paired T-Test comparing the temporal encoding approaches to check for statistical significance between pairs
out of our initial hypothesis and detect a significant difference between JP and ANC. In terms of accuracy, we discover a significant difference between SI (62.1\%) and AN (68.6\%). Whereas, between SI and JP (64.45\%) or ANC (59.13\%) there is no significant difference. We conjecture these results to be due to the graph's size and limited number of structural changes over time, that might favor AN as it is possible for participants to follow all changes the during animation. Our analysis shows no evidence to support H2.

\begin{figure}[!h]
     \centering
     \includegraphics[width=\textwidth]{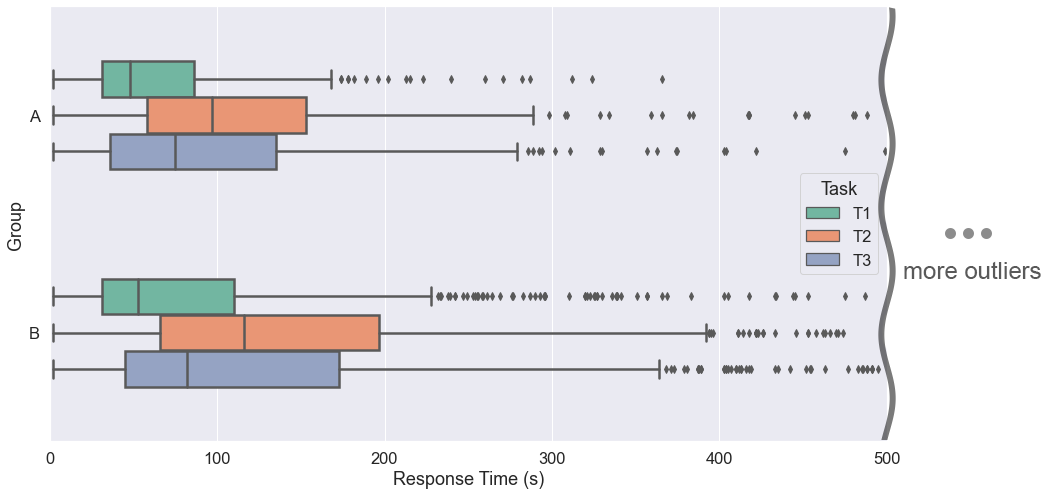}
     \caption{\textbf{H3:} Box plot of response times for interaction groups per task.}
     \label{fig:h3_response}
\end{figure}

\noindent\textbf{H3.} 
We conjecture that providing interactions
influences the response times but not the accuracy. Our tests detect a significant difference (see Table~\ref{tab:results}) in the response times between group A (no interactions; 114.76s) and group B (interactions; 163.83s). As we initially assume, the group with interactions is much slower in completing tasks than the group with no interactions (see Figure~\ref{fig:h3_response}), however, the difference in accuracy is unexpected. The group with interactions is significantly more accurate than the one without (group A: 58\%, group B: 65\%). This suggests that interactions indeed increase response times, but at the same time provide the participants with a much better understanding of the visualized graphs and corresponding network dynamics regardless of the temporal encoding, therefore, leading to more accurate responses. The analysis shows that our results support H3 
in terms of response times, but not accuracy.

\begin{figure}[!h]
     \centering
     \includegraphics[width=\textwidth]{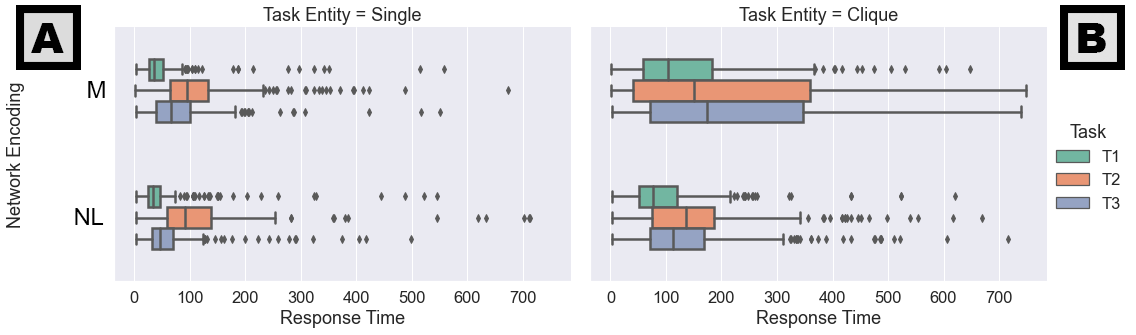}
     \caption{\textbf{H4:} Box plot of response times for (A) single entities and (B) cliques.}
     \label{fig:h4_response}
\end{figure}

\noindent\textbf{H4.} 
We formulate this hypothesis to evaluate whether the response times and accuracy of M and NL representations is affected by the type of target entity in a dynamic context (\textit{low-level} - individual nodes and links; or \textit{higher-level} - cliques), regardless of the temporal representation.
For low-level entities, we do not detect any significant differences of the response times between network representations (see Table~\ref{tab:results}), both NL and M diagrams perform similarly with no clear winner. The results (see Figure~\ref{fig:h4_response}) for tasks on low-level entities indicate that M has lower response times (NL: 97.08s , M: 90.24s), whereas for higher-level entities NL has significantly lower response times (NL: 146.66s, M: 245.2s). However, in terms of accuracy M is significantly better than NL for lower-level entities (NL: 82.1\%, M: 86.4\%). For the higher-level entities, NL and M representations perform quite similarly in terms of accuracy (NL: 42.1\%, M: 41.3\%) Based on these findings, the results suggest that H4 is partially supported.

\begin{figure}[h]
     \centering
     \includegraphics[width=\textwidth]{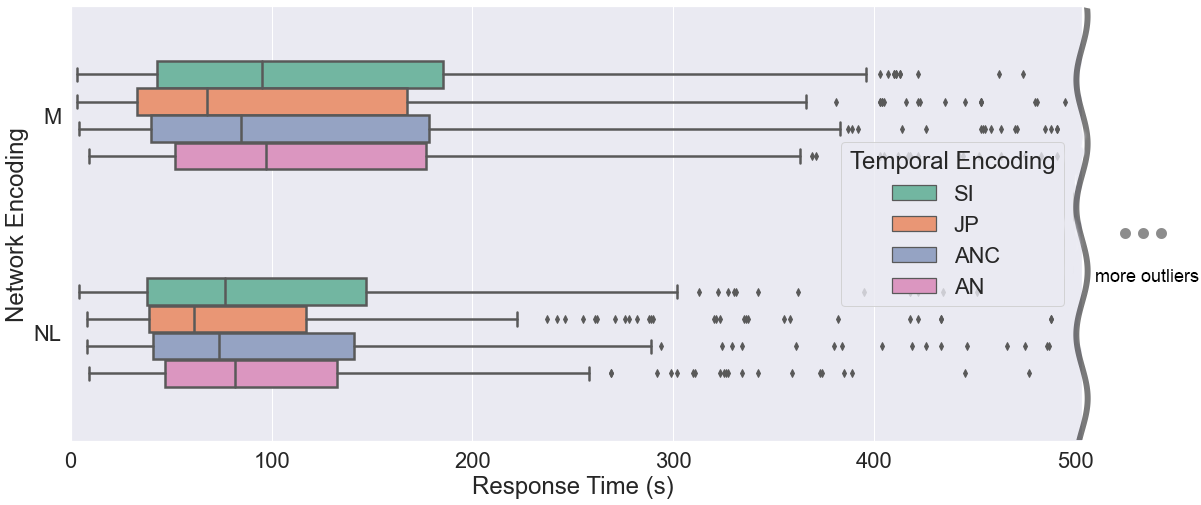}
         \caption{\textbf{H5:} Box plot of response times for temporal and network representations.}
     \label{fig:h5_response}
\end{figure}

\noindent\textbf{H5.} 
Finally, we want to assess the response times and accuracy for all possible combinations of network structural and temporal encodings. Our assumption is that M representations with SI temporal encoding have the lowest response times and highest accuracy. We compare M+SI to all other combinations of network structural and temporal encodings (see Figure~\ref{fig:h5_response}). 
The results of the statistical tests yield significant differences in response times when comparing M+SI (154.53s) with M+JP (140.13s), NL+SI (105.25s), NL+JP (99.54s), NL+AN (108.8s), and NL+ANC (110.97s). Between M+SI (154.53s) and M+ANC (168.87s) and M+AN (160.62s) there is no significant difference in response times (see Table~\ref{tab:results}). In terms of accuracy we detect statistically significant differences between M+SI (61.1\%) and NL+ANC (51.8\%) and NL+AN (71.4\%). Whereas, the other combinations do not differ enough to warrant significance: M+JP (64\%), M+ANC (66.4\%), M+AN (65.5\%), NL+JP (64.6\%), and NL+SI (62.9\%). From these results, the most balanced combination in terms of response times and accuracy is NL+AN followed by NL+JP. Therefore, we find no evidence supporting H5. 

\subsection{Qualitative Results}
\begin{figure}[t]
     \centering
     \includegraphics[width=0.9\textwidth]{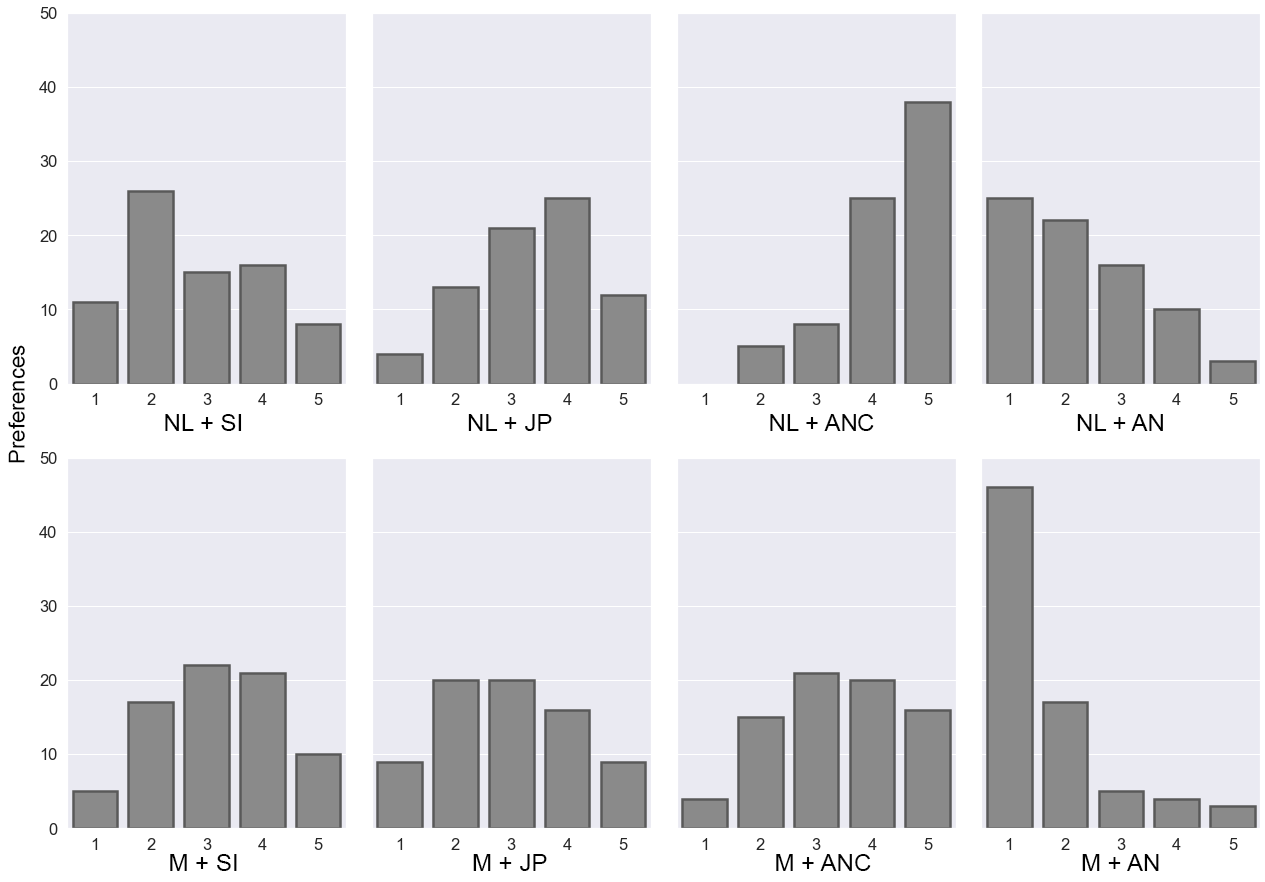}
     \caption{Preferences per network and temporal encoding on a Likert scale (1-5).}
     \label{fig:preferences}
\end{figure}

\begin{figure}[t]
     \centering
     \includegraphics[width=0.9\textwidth]{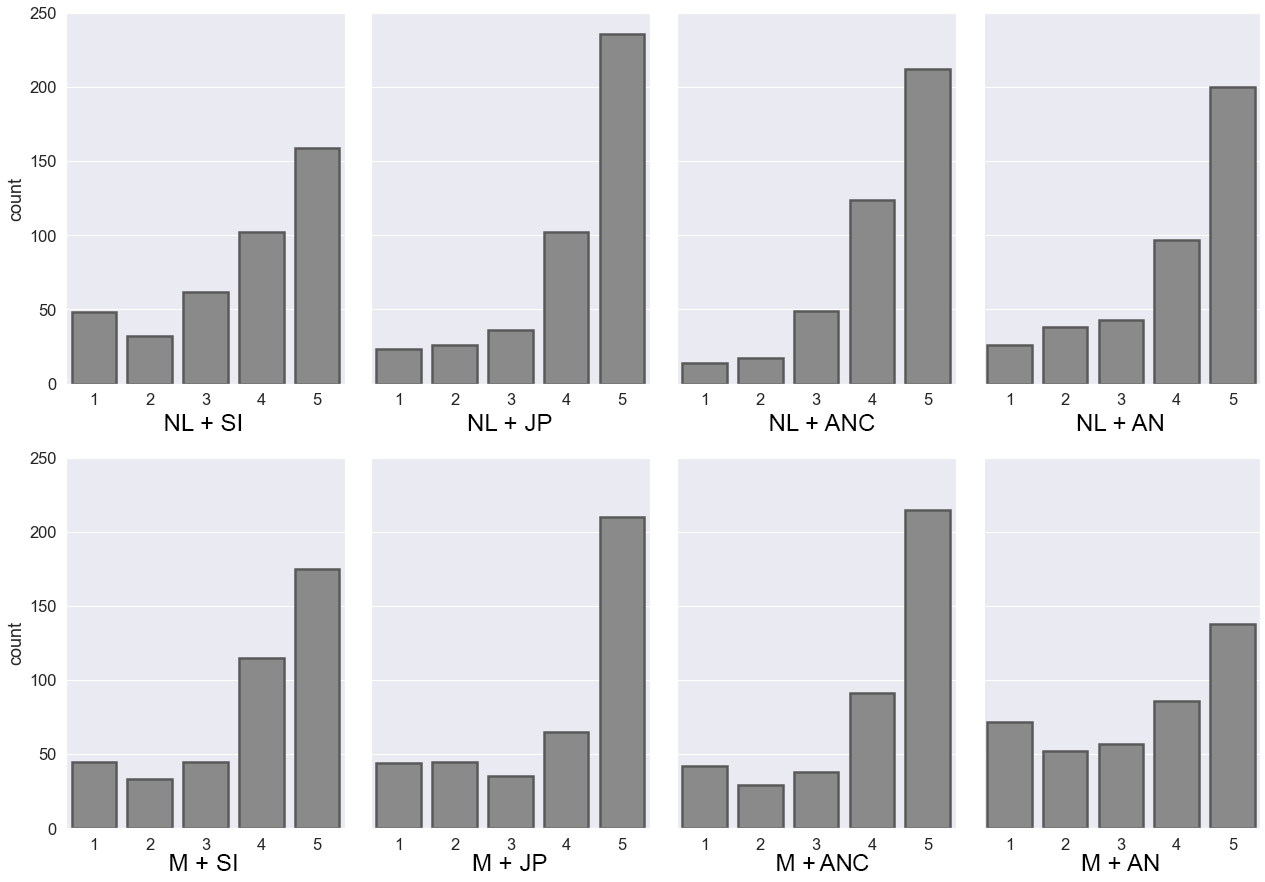}
     \caption{Confidence per network and temporal encoding on a Likert scale (1-5).}
     \label{fig:confidence}
\end{figure}

We collect the participants' ratings per combination of network structural and temporal encoding along with textual feedback pertaining to their preferences and experience during the experiment (see Figure~\ref{fig:preferences}). There are no major differences in the preferences between the SI and JP encodings; ANC is the most preferred temporal encoding when coupled with a NL base representation. 
The NL representation is generally the most preferred approach, regardless of the temporal encoding. In terms of the participants confidence,
we observe that most participants seemed to be fairly confident in their answers across all approaches (see Figure~\ref{fig:confidence}). Most notably, the participants were most confident with NL+JP, followed by M+ANC, M+JP, and NL+ANC. 
There is general consensus that NL+SI was a very cluttered combination, whereas for M it performed a lot better and was easier to understand (``\textit{SI was really confusing for some of the NL tasks but really useful for many of the M tasks}''). This is presumably due to the clutter generated by parallel edges crossings that occur in NL diagrams, which does not affect M. 
As in previous studies~\cite{filipov2021exploratory}, 
the feedback on JP 
outlines that it requires participants to split their attention between multiple views in order to compare the temporal information. 
The ANC approach was 
preferred by the study participants for its 
flexibility 
due to the additional controls (i.e., time slider).
AN was not considered to be a very good temporal encoding technique with the feedback being consistent across structural representations. Some participants commented that they needed to ``\textit{screenshot every timestamp to look at the different connections between the nodes}'' and wait to watch the whole animation from the beginning. NL+AN,  therefore, appears to be the least practical of the approaches, however, it also provides the best results. We conjecture this to be due to the size of the graphs and the amount of structural changes occurring. M+AN is the lowest rated by the participants. The general consensus for AN is that it was difficult to keep track of the changes occurring between the nodes, requiring the viewer to memorize node positions and labels incurring a high cognitive effort to complete the tasks. Despite the aforementioned drawbacks, AN scales better to a larger amount of timeslices compared to SI and JP.
Finally, the group with interactions had a better experience overall compared to the group without. The majority of the members of this group explicitly requested interactions to be implemented, supporting our findings concerning H3.


\subsection{Limitations}

This experiment's limitations open potential future research directions. First, the \textbf{size} of the graph was not considered. We chose small graphs as stimuli for this study, both in the amount of nodes/links and number of timeslices. 
M scales better to larger graphs than NL, 
while AN and ANC 
support a greater number of timeslices compared to SI and JP. Future studies on dynamic network visualizations might provide evidence on the scalability of the different potential combinations.
Second, we chose simple, custom implementations for our structural and temporal encodings, disregarding more advanced solutions in literature (see Section~\ref{sec:design_space}). While this was done with the intention of testing the fundamental principles of the techniques in our design space, evaluating more sophisticated approaches might have significantly impacted the results. Finally, we focus on a selection of tasks
from a taxonomy on network evolution~\cite{ahn2014}, other graph-based taxonomies could present 
relevent benchmarks for the proposed techniques.
 
\section{Conclusion}
\label{sec:conclusion}
In this paper we presented an experimental study assessing the response times, accuracy, and preferences of participants on different combinations of network structural and temporal encodings, with and without interaction support for the network exploration. Overall, the participants expressed a preference for NL over M, specifically preferring the ANC temporal encoding over the other options, despite AN being more accurate and having lower response times. We also note that our results suggest that the use of M as base representation 
proved to be more accurate for tasks on low-level entities and counting across different temporal representations.
The results of our experiment also suggest a significant effect of interactions on participants' performance. Therefore, as directions for future work, we consider evaluating in more detail the influence that interactions have on accuracy and response times for dynamic network visualization, also considering the potential influence of the graph size on the perception of different combinations of network and temporal encodings.

%
%

\noindent \textbf{Acknowledgements} This work was conducted within the framework of the project KnoVA (P31419-N31) and ArtVis (P35767) funded by the Austrian Science Fund (FWF).
\bibliographystyle{splncs04}
\bibliography{bib}

\setcounter{section}{0}
\setcounter{figure}{0}    
\setcounter{table}{0}    

\section*{Appendix}

In this Appendix we include supplemental results which complement our discussion in the paper. The study and downloadable material is accessible online:  \href{https://graph-survey.herokuapp.com/}{https://graph-survey.herokuapp.com/}. 

The study initially received 76 submissions. We introduced a control question every 12 the filter out participants trying to game the study, resulting in 8 being removed.

\begin{table}[h]
    \centering
    \begin{tabular}{c|c|c}
         & Low-level & High-level \\ \hline
         \textbf{T1} & \begin{minipage}[t]{0.42\linewidth}
         At which time step is the relationship between \{source\} and \{target\} introduced for the first time?  
        \end{minipage} & \begin{minipage}[t]{0.42\linewidth}
        At which time step does the clique between \{nodes\} appear for the first time?
        \end{minipage}  \\ \hline
        \textbf{T2} & \begin{minipage}[t]{0.42\linewidth}
         Sum up the  changes (additions and removals) of \{node\}’s degree across all time steps.  
        \end{minipage} & \begin{minipage}[t]{0.42\linewidth}
        Calculate change of the clique’s size between \{nodes\} across all time steps.
        \end{minipage}  \\ \hline
        \textbf{T3} & \begin{minipage}[t]{0.42\linewidth}
          At which time step does the node \{node\} have its highest degree?  
        \end{minipage} & \begin{minipage}[t]{0.42\linewidth}
       Consider the set of nodes \{nodes\}. Find the size of the largest maximal clique across all the time steps between the given nodes.
        \end{minipage}  \\ \hline
    \end{tabular}
    \caption{Example questions, per task (rows) and entity type (columns). Target nodes for low-level tasks were chosen randomly during the study design; cliques were also introduced artificially as explained in Section~\ref{sec:eval_setting} of the paper.}
    \label{tab:appendix_example_questions}
\end{table}

\begin{figure}
    \centering
    \begin{subfigure}[c]{\textwidth}
        \includegraphics[width=\textwidth]{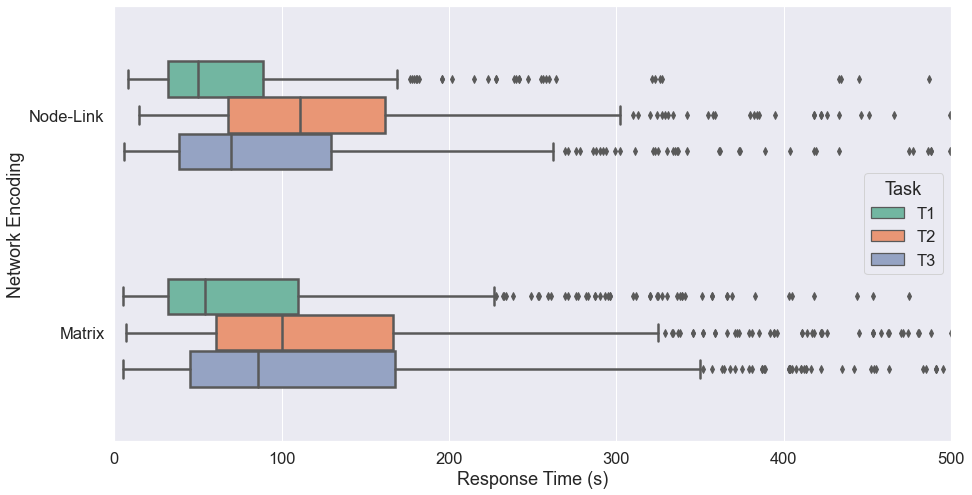}
        \subcaption{Differences in response times between node-link and matrix network encodings per task.}
    \end{subfigure}
        \begin{subfigure}[c]{\textwidth}
        \includegraphics[width=\textwidth]{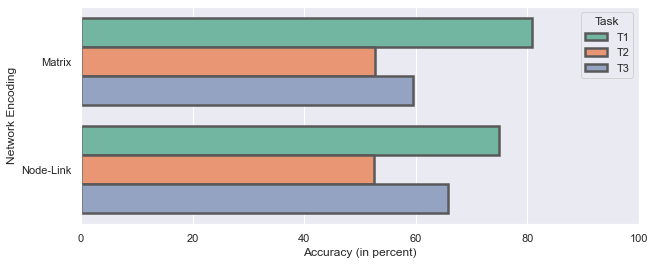}
        \subcaption{Accuracy (in percent) for node-link and matrix representations per task.}
    \end{subfigure}
    \caption{Response times and correct responses between network encodings and tasks for \textbf{H1: Matrices have lower response times and higher accuracy for all tasks compared to node-link diagrams, regardless of the temporal encoding.                                         }}
\clearpage

\end{figure}


\begin{figure}
    \centering
    \begin{subfigure}[c]{\textwidth}
        \includegraphics[width=\textwidth]{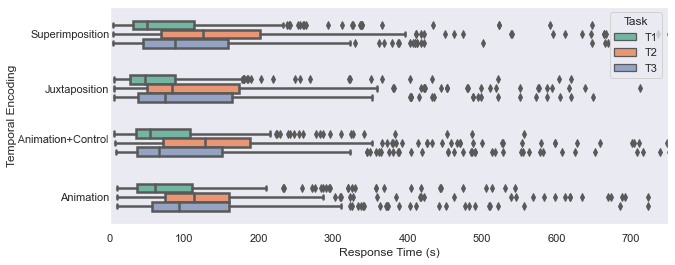}
        \subcaption{Differences in response times between temporal encodings per task.}
    \end{subfigure}
        \begin{subfigure}[c]{\textwidth}
        \includegraphics[width=\textwidth]{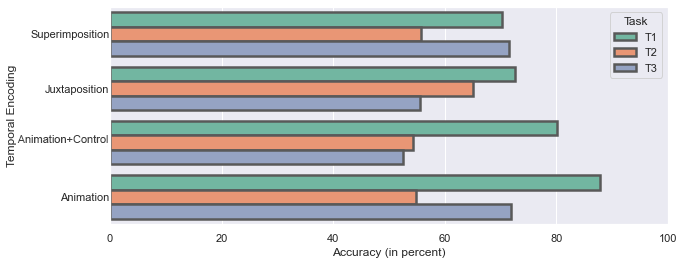}
        \subcaption{Accuracy (in percent) for temporal encodings per task.}
    \end{subfigure}
    \caption{Response times and accuracy between network encodings and tasks for \textbf{H2: From all temporal encoding techniques, superimposition has the lowest response times and highest accuracy, regardless of the structural representation.               }}
\end{figure}


\begin{figure}
    \centering
    \begin{subfigure}[c]{\textwidth}
        \includegraphics[width=\textwidth]{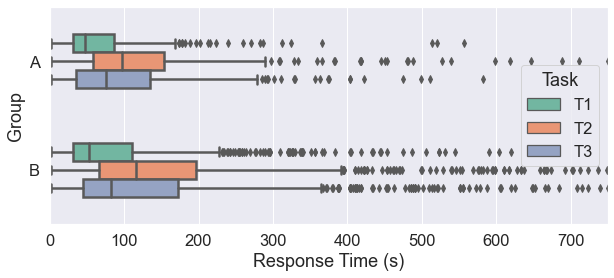}
        \subcaption{Differences in response times between interaction groups per task.}
    \end{subfigure}
        \begin{subfigure}[c]{\textwidth}
        \includegraphics[width=\textwidth]{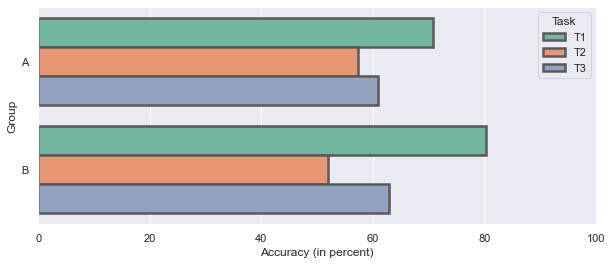}
        \subcaption{Accuracy (in percent) for interaction groups per task.}
    \end{subfigure}
    \caption{Response times and accuracy between groups (A - no interaction; B - interaction) and tasks for \textbf{H3: Providing interaction techniques increases the response times but not the accuracy.}}
\end{figure}


\begin{figure}
    \centering
    \begin{subfigure}[c]{\textwidth}
        \includegraphics[width=\textwidth]{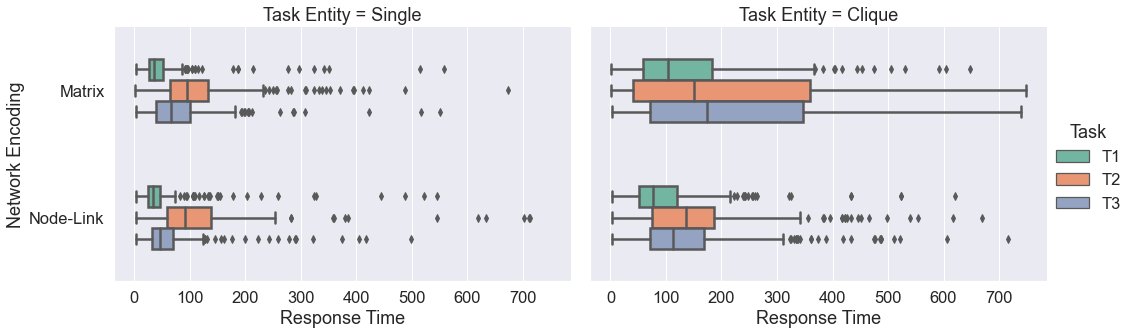}
        \subcaption{Differences in response times between network encodings and entity type per task.}
    \end{subfigure}
    \begin{subfigure}[c]{\textwidth}
        \includegraphics[width=\textwidth]{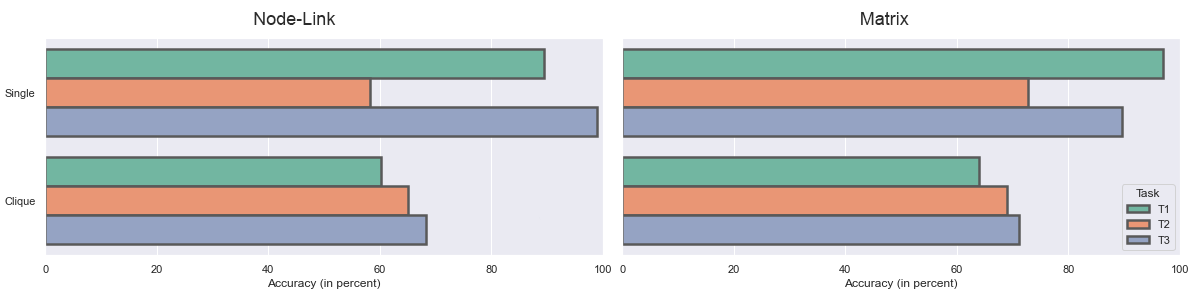}
        \subcaption{Accuracy (in percent) for adjacency matrices and entity type per task.}
    \end{subfigure}
    \caption{Response times and accuracy between network representations and entity types (single - low-level; clique - high-level) and tasks for \textbf{H4:Matrices have lower response times and higher accuracy for tasks on low-level entities and node-link diagrams have lower response times and higher accuracy for tasks on higher-level entities, regardless oft he temporal encoding. Matrices have lower response times and higher accuracy for tasks on low-level entities and node-link diagrams have lower response times and higher accuracy for tasks on higher-level entities, regardless oft he temporal encoding.}}
\end{figure}

\begin{figure}
    \centering
    \begin{subfigure}[c]{\textwidth}
        \includegraphics[width=\textwidth]{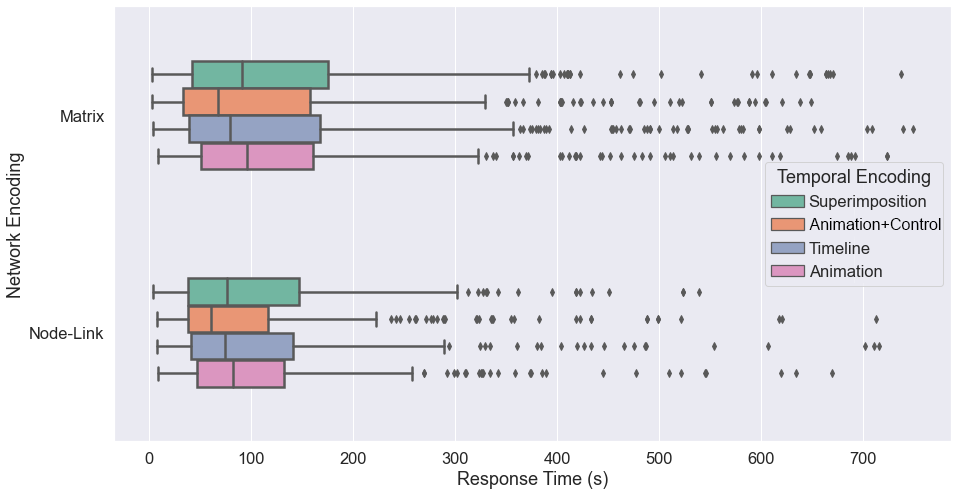}
        \subcaption{Differences in response times between network representations and temporal encodings.}
    \end{subfigure}
        \begin{subfigure}[c]{\textwidth}
        \includegraphics[width=\textwidth]{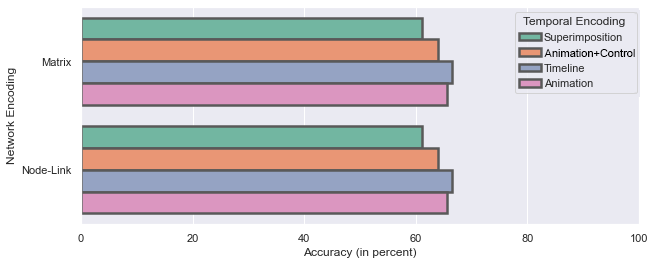}
        \subcaption{Accuracy (in percent) for different combinations of network representations and temporal encodings.}
    \end{subfigure}
    \caption{Response times and accuracy between network representations and temporal encodings for \textbf{H5: The combination of matrices with superimposition results in the lowest response times and highest accuracy compared to other combinations of network structural and temporal encoding.}}
\end{figure}



\newpage
\clearpage

In the following we provide the results of our Analysis of Variance (ANOVA tests for each hypothesis investigating the influence of the independent variables in our study (task type, network and temporal encoding, and group) on the dependent variable time, outlining the significance levels for each. Column ``Df'' refers to the degrees of freedom for the independent variable (number of levels in the variable minus 1); ``Sum Sq'' displays the sum of squares (a.k.a. the total variation between the group means and the overall mean); ``Mean Sq'' is the mean of the sum of the squares, calculated by dividing the sum of squares by the degrees of freedom for each parameter; ``F Value'' column is the test statistic from the F-test. It is the mean square of each independent variable divided by the mean square of the residuals: the larger the value, the more likely it is that the variation caused by the independent variable is real and not due to chance. Finally, ``Pr($>$F)'' is the \textit{p}-value of the F-test statistics,  and the  ``Significance'' column reports a symbol for quick visual reference about statistical significance.


\begin{table}[h]
\begin{tabular}{|
>{\columncolor[HTML]{FFFFFF}}l 
>{\columncolor[HTML]{FFFFFF}}r 
>{\columncolor[HTML]{FFFFFF}}r 
>{\columncolor[HTML]{FFFFFF}}r 
>{\columncolor[HTML]{FFFFFF}}r 
>{\columncolor[HTML]{FFFFFF}}r 
>{\columncolor[HTML]{FFFFFF}}l |}
\hline
\multicolumn{1}{|l|}{\cellcolor[HTML]{FFFFFF}\textbf{Response Time}} &
  \multicolumn{1}{l|}{\cellcolor[HTML]{FFFFFF}Df} &
  \multicolumn{1}{l|}{\cellcolor[HTML]{FFFFFF}Sum Sq} &
  \multicolumn{1}{l|}{\cellcolor[HTML]{FFFFFF}Mean Sq} &
  \multicolumn{1}{l|}{\cellcolor[HTML]{FFFFFF}F value} &
  \multicolumn{1}{l|}{\cellcolor[HTML]{FFFFFF}Pr(\textgreater{}F)} &
  Significance \\ \hline
\multicolumn{1}{|l|}{\cellcolor[HTML]{FFFFFF}network\_enc} &
  \multicolumn{1}{r|}{\cellcolor[HTML]{FFFFFF}1} &
  \multicolumn{1}{r|}{\cellcolor[HTML]{FFFFFF}2138106} &
  \multicolumn{1}{r|}{\cellcolor[HTML]{FFFFFF}2138106} &
  \multicolumn{1}{r|}{\cellcolor[HTML]{FFFFFF}98.0501} &
  \multicolumn{1}{r|}{\cellcolor[HTML]{FFFFFF}\textless 2.2e-16} &
  \multicolumn{1}{r|}{\cellcolor[HTML]{FFFFFF}***} \\ \hline
\multicolumn{1}{|l|}{\cellcolor[HTML]{FFFFFF}task\_type} &
  \multicolumn{1}{r|}{\cellcolor[HTML]{FFFFFF}2} &
  \multicolumn{1}{r|}{\cellcolor[HTML]{FFFFFF}3523697} &
  \multicolumn{1}{r|}{\cellcolor[HTML]{FFFFFF}1761849} &
  \multicolumn{1}{r|}{\cellcolor[HTML]{FFFFFF}80.7955} &
  \multicolumn{1}{r|}{\cellcolor[HTML]{FFFFFF}\textless 2.2e-16} &
  \multicolumn{1}{r|}{\cellcolor[HTML]{FFFFFF}***} \\ \hline
\multicolumn{1}{|l|}{\cellcolor[HTML]{FFFFFF}group} &
  \multicolumn{1}{r|}{\cellcolor[HTML]{FFFFFF}1} &
  \multicolumn{1}{r|}{\cellcolor[HTML]{FFFFFF}338934} &
  \multicolumn{1}{r|}{\cellcolor[HTML]{FFFFFF}338934} &
  \multicolumn{1}{r|}{\cellcolor[HTML]{FFFFFF}15.543} &
  \multicolumn{1}{r|}{\cellcolor[HTML]{FFFFFF}8.24E-05} &
  \multicolumn{1}{r|}{\cellcolor[HTML]{FFFFFF}***} \\ \hline
\multicolumn{1}{|l|}{\cellcolor[HTML]{FFFFFF}network\_enc:task\_type} &
  \multicolumn{1}{r|}{\cellcolor[HTML]{FFFFFF}2} &
  \multicolumn{1}{r|}{\cellcolor[HTML]{FFFFFF}303223} &
  \multicolumn{1}{r|}{\cellcolor[HTML]{FFFFFF}151612} &
  \multicolumn{1}{r|}{\cellcolor[HTML]{FFFFFF}6.9527} &
  \multicolumn{1}{r|}{\cellcolor[HTML]{FFFFFF}0.0009704} &
  \multicolumn{1}{r|}{\cellcolor[HTML]{FFFFFF}***} \\ \hline
\multicolumn{1}{|l|}{\cellcolor[HTML]{FFFFFF}network\_enc:group} &
  \multicolumn{1}{r|}{\cellcolor[HTML]{FFFFFF}1} &
  \multicolumn{1}{r|}{\cellcolor[HTML]{FFFFFF}30489} &
  \multicolumn{1}{r|}{\cellcolor[HTML]{FFFFFF}30489} &
  \multicolumn{1}{r|}{\cellcolor[HTML]{FFFFFF}1.3982} &
  \multicolumn{1}{r|}{\cellcolor[HTML]{FFFFFF}0.2371122} &
   \\ \hline
\multicolumn{1}{|l|}{\cellcolor[HTML]{FFFFFF}task\_type:group} &
  \multicolumn{1}{r|}{\cellcolor[HTML]{FFFFFF}2} &
  \multicolumn{1}{r|}{\cellcolor[HTML]{FFFFFF}26350} &
  \multicolumn{1}{r|}{\cellcolor[HTML]{FFFFFF}13175} &
  \multicolumn{1}{r|}{\cellcolor[HTML]{FFFFFF}0.6042} &
  \multicolumn{1}{r|}{\cellcolor[HTML]{FFFFFF}0.5465832} &
   \\ \hline
\multicolumn{1}{|l|}{\cellcolor[HTML]{FFFFFF}network\_enc:task\_type:group} &
  \multicolumn{1}{r|}{\cellcolor[HTML]{FFFFFF}2} &
  \multicolumn{1}{r|}{\cellcolor[HTML]{FFFFFF}9345} &
  \multicolumn{1}{r|}{\cellcolor[HTML]{FFFFFF}4672} &
  \multicolumn{1}{r|}{\cellcolor[HTML]{FFFFFF}0.2143} &
  \multicolumn{1}{r|}{\cellcolor[HTML]{FFFFFF}0.807143} &
   \\ \hline
\multicolumn{1}{|l|}{\cellcolor[HTML]{FFFFFF}Residuals} &
  \multicolumn{1}{r|}{\cellcolor[HTML]{FFFFFF}3250} &
  \multicolumn{1}{r|}{\cellcolor[HTML]{FFFFFF}70870339} &
  \multicolumn{1}{r|}{\cellcolor[HTML]{FFFFFF}21806} &
  \multicolumn{1}{l|}{\cellcolor[HTML]{FFFFFF}} &
  \multicolumn{1}{l|}{\cellcolor[HTML]{FFFFFF}} &
   \\ \hline
\multicolumn{7}{|l|}{\cellcolor[HTML]{FFFFFF}Signif. codes:  0 ‘***’ 0.001 ‘**’ 0.01 ‘*’ 0.05 ‘.’ 0.1 ‘ ’ 1} \\ \hline
\end{tabular}
\caption{ANOVA table for \textbf{H1} showing the influence of the independent variables on the dependent variable ``time'' together with their interaction effects.}
\end{table}


\begin{table}[]
\begin{tabular}{|
>{\columncolor[HTML]{FFFFFF}}l 
>{\columncolor[HTML]{FFFFFF}}r 
>{\columncolor[HTML]{FFFFFF}}r 
>{\columncolor[HTML]{FFFFFF}}r 
>{\columncolor[HTML]{FFFFFF}}r 
>{\columncolor[HTML]{FFFFFF}}r 
>{\columncolor[HTML]{FFFFFF}}l |}
\hline
\multicolumn{1}{|l|}{\cellcolor[HTML]{FFFFFF}\textbf{Response Time}} &
  \multicolumn{1}{l|}{\cellcolor[HTML]{FFFFFF}Df} &
  \multicolumn{1}{l|}{\cellcolor[HTML]{FFFFFF}Sum Sq} &
  \multicolumn{1}{l|}{\cellcolor[HTML]{FFFFFF}Mean Sq} &
  \multicolumn{1}{l|}{\cellcolor[HTML]{FFFFFF}F value} &
  \multicolumn{1}{l|}{\cellcolor[HTML]{FFFFFF}Pr(\textgreater{}F)} &
  Significance \\ \hline
\multicolumn{1}{|l|}{\cellcolor[HTML]{FFFFFF}network\_enc} &
  \multicolumn{1}{r|}{\cellcolor[HTML]{FFFFFF}1} &
  \multicolumn{1}{r|}{\cellcolor[HTML]{FFFFFF}1357060} &
  \multicolumn{1}{r|}{\cellcolor[HTML]{FFFFFF}1357060} &
  \multicolumn{1}{r|}{\cellcolor[HTML]{FFFFFF}61.7501} &
  \multicolumn{1}{r|}{\cellcolor[HTML]{FFFFFF}5.25E-15} &
  \multicolumn{1}{r|}{\cellcolor[HTML]{FFFFFF}***} \\ \hline
\multicolumn{1}{|l|}{\cellcolor[HTML]{FFFFFF}task\_type} &
  \multicolumn{1}{r|}{\cellcolor[HTML]{FFFFFF}3} &
  \multicolumn{1}{r|}{\cellcolor[HTML]{FFFFFF}213354} &
  \multicolumn{1}{r|}{\cellcolor[HTML]{FFFFFF}71118} &
  \multicolumn{1}{r|}{\cellcolor[HTML]{FFFFFF}3.2361} &
  \multicolumn{1}{r|}{\cellcolor[HTML]{FFFFFF}0.02134} &
  \multicolumn{1}{r|}{\cellcolor[HTML]{FFFFFF}*} \\ \hline
\multicolumn{1}{|l|}{\cellcolor[HTML]{FFFFFF}group} &
  \multicolumn{1}{r|}{\cellcolor[HTML]{FFFFFF}1} &
  \multicolumn{1}{r|}{\cellcolor[HTML]{FFFFFF}346104} &
  \multicolumn{1}{r|}{\cellcolor[HTML]{FFFFFF}346104} &
  \multicolumn{1}{r|}{\cellcolor[HTML]{FFFFFF}15.7487} &
  \multicolumn{1}{r|}{\cellcolor[HTML]{FFFFFF}7.39E-05} &
  \multicolumn{1}{r|}{\cellcolor[HTML]{FFFFFF}***} \\ \hline
\multicolumn{1}{|l|}{\cellcolor[HTML]{FFFFFF}network\_enc:task\_type} &
  \multicolumn{1}{r|}{\cellcolor[HTML]{FFFFFF}3} &
  \multicolumn{1}{r|}{\cellcolor[HTML]{FFFFFF}75322} &
  \multicolumn{1}{r|}{\cellcolor[HTML]{FFFFFF}25107} &
  \multicolumn{1}{r|}{\cellcolor[HTML]{FFFFFF}1.1425} &
  \multicolumn{1}{r|}{\cellcolor[HTML]{FFFFFF}0.33047} &
  \multicolumn{1}{r|}{\cellcolor[HTML]{FFFFFF}} \\ \hline
\multicolumn{1}{|l|}{\cellcolor[HTML]{FFFFFF}network\_enc:group} &
  \multicolumn{1}{r|}{\cellcolor[HTML]{FFFFFF}1} &
  \multicolumn{1}{r|}{\cellcolor[HTML]{FFFFFF}2920} &
  \multicolumn{1}{r|}{\cellcolor[HTML]{FFFFFF}2920} &
  \multicolumn{1}{r|}{\cellcolor[HTML]{FFFFFF}0.1329} &
  \multicolumn{1}{r|}{\cellcolor[HTML]{FFFFFF}0.7155} &
   \\ \hline
\multicolumn{1}{|l|}{\cellcolor[HTML]{FFFFFF}task\_type:group} &
  \multicolumn{1}{r|}{\cellcolor[HTML]{FFFFFF}3} &
  \multicolumn{1}{r|}{\cellcolor[HTML]{FFFFFF}5082} &
  \multicolumn{1}{r|}{\cellcolor[HTML]{FFFFFF}1694} &
  \multicolumn{1}{r|}{\cellcolor[HTML]{FFFFFF}0.0771} &
  \multicolumn{1}{r|}{\cellcolor[HTML]{FFFFFF}0.97239} &
   \\ \hline
\multicolumn{1}{|l|}{\cellcolor[HTML]{FFFFFF}network\_enc:task\_type:group} &
  \multicolumn{1}{r|}{\cellcolor[HTML]{FFFFFF}3} &
  \multicolumn{1}{r|}{\cellcolor[HTML]{FFFFFF}36402} &
  \multicolumn{1}{r|}{\cellcolor[HTML]{FFFFFF}12134} &
  \multicolumn{1}{r|}{\cellcolor[HTML]{FFFFFF}0.5521} &
  \multicolumn{1}{r|}{\cellcolor[HTML]{FFFFFF}0.64671} &
   \\ \hline
\multicolumn{1}{|l|}{\cellcolor[HTML]{FFFFFF}Residuals} &
  \multicolumn{1}{r|}{\cellcolor[HTML]{FFFFFF}3262} &
  \multicolumn{1}{r|}{\cellcolor[HTML]{FFFFFF}71687780} &
  \multicolumn{1}{r|}{\cellcolor[HTML]{FFFFFF}21977} &
  \multicolumn{1}{l|}{\cellcolor[HTML]{FFFFFF}} &
  \multicolumn{1}{l|}{\cellcolor[HTML]{FFFFFF}} &
   \\ \hline
\multicolumn{7}{|l|}{\cellcolor[HTML]{FFFFFF}Signif. codes:  0 ‘***’ 0.001 ‘**’ 0.01 ‘*’ 0.05 ‘.’ 0.1 ‘ ’ 1} \\ \hline
\end{tabular}
\caption{ANOVA table for \textbf{H2} showing the influence of the independent variables on the dependent variable ``time'' together with their interaction effects.}
\end{table}



\begin{table}[]
\begin{tabular}{|
>{\columncolor[HTML]{FFFFFF}}l 
>{\columncolor[HTML]{FFFFFF}}r 
>{\columncolor[HTML]{FFFFFF}}r 
>{\columncolor[HTML]{FFFFFF}}r 
>{\columncolor[HTML]{FFFFFF}}r 
>{\columncolor[HTML]{FFFFFF}}r 
>{\columncolor[HTML]{FFFFFF}}l |}
\hline
\multicolumn{1}{|l|}{\cellcolor[HTML]{FFFFFF}\textbf{Response Time}} &
  \multicolumn{1}{l|}{\cellcolor[HTML]{FFFFFF}Df} &
  \multicolumn{1}{l|}{\cellcolor[HTML]{FFFFFF}Sum Sq} &
  \multicolumn{1}{l|}{\cellcolor[HTML]{FFFFFF}Mean Sq} &
  \multicolumn{1}{l|}{\cellcolor[HTML]{FFFFFF}F value} &
  \multicolumn{1}{l|}{\cellcolor[HTML]{FFFFFF}Pr(\textgreater{}F)} &
  Significance \\ \hline
\multicolumn{1}{|l|}{\cellcolor[HTML]{FFFFFF}network\_enc} &
  \multicolumn{1}{r|}{\cellcolor[HTML]{FFFFFF}1} &
  \multicolumn{1}{r|}{\cellcolor[HTML]{FFFFFF}867172} &
  \multicolumn{1}{r|}{\cellcolor[HTML]{FFFFFF}867172} &
  \multicolumn{1}{r|}{\cellcolor[HTML]{FFFFFF}9.1825} &
  \multicolumn{1}{r|}{\cellcolor[HTML]{FFFFFF}2.46E-03} &
  \multicolumn{1}{r|}{\cellcolor[HTML]{FFFFFF}**} \\ \hline
\multicolumn{1}{|l|}{\cellcolor[HTML]{FFFFFF}task\_type} &
  \multicolumn{1}{r|}{\cellcolor[HTML]{FFFFFF}3} &
  \multicolumn{1}{r|}{\cellcolor[HTML]{FFFFFF}219958} &
  \multicolumn{1}{r|}{\cellcolor[HTML]{FFFFFF}73319} &
  \multicolumn{1}{r|}{\cellcolor[HTML]{FFFFFF}0.7764} &
  \multicolumn{1}{r|}{\cellcolor[HTML]{FFFFFF}0.507051} &
  \multicolumn{1}{r|}{\cellcolor[HTML]{FFFFFF}} \\ \hline
\multicolumn{1}{|l|}{\cellcolor[HTML]{FFFFFF}group} &
  \multicolumn{1}{r|}{\cellcolor[HTML]{FFFFFF}1} &
  \multicolumn{1}{r|}{\cellcolor[HTML]{FFFFFF}1537636} &
  \multicolumn{1}{r|}{\cellcolor[HTML]{FFFFFF}1537636} &
  \multicolumn{1}{r|}{\cellcolor[HTML]{FFFFFF}16.282} &
  \multicolumn{1}{r|}{\cellcolor[HTML]{FFFFFF}5.58E-05} &
  \multicolumn{1}{r|}{\cellcolor[HTML]{FFFFFF}***} \\ \hline
\multicolumn{1}{|l|}{\cellcolor[HTML]{FFFFFF}network\_enc:task\_type} &
  \multicolumn{1}{r|}{\cellcolor[HTML]{FFFFFF}3} &
  \multicolumn{1}{r|}{\cellcolor[HTML]{FFFFFF}65818} &
  \multicolumn{1}{r|}{\cellcolor[HTML]{FFFFFF}21939} &
  \multicolumn{1}{r|}{\cellcolor[HTML]{FFFFFF}0.2323} &
  \multicolumn{1}{r|}{\cellcolor[HTML]{FFFFFF}0.873916} &
  \multicolumn{1}{r|}{\cellcolor[HTML]{FFFFFF}} \\ \hline
\multicolumn{1}{|l|}{\cellcolor[HTML]{FFFFFF}network\_enc:group} &
  \multicolumn{1}{r|}{\cellcolor[HTML]{FFFFFF}1} &
  \multicolumn{1}{r|}{\cellcolor[HTML]{FFFFFF}255} &
  \multicolumn{1}{r|}{\cellcolor[HTML]{FFFFFF}255} &
  \multicolumn{1}{r|}{\cellcolor[HTML]{FFFFFF}0.0027} &
  \multicolumn{1}{r|}{\cellcolor[HTML]{FFFFFF}0.958563} &
   \\ \hline
\multicolumn{1}{|l|}{\cellcolor[HTML]{FFFFFF}task\_type:group} &
  \multicolumn{1}{r|}{\cellcolor[HTML]{FFFFFF}3} &
  \multicolumn{1}{r|}{\cellcolor[HTML]{FFFFFF}44034} &
  \multicolumn{1}{r|}{\cellcolor[HTML]{FFFFFF}14678} &
  \multicolumn{1}{r|}{\cellcolor[HTML]{FFFFFF}0.1554} &
  \multicolumn{1}{r|}{\cellcolor[HTML]{FFFFFF}0.926229} &
   \\ \hline
\multicolumn{1}{|l|}{\cellcolor[HTML]{FFFFFF}network\_enc:task\_type:group} &
  \multicolumn{1}{r|}{\cellcolor[HTML]{FFFFFF}3} &
  \multicolumn{1}{r|}{\cellcolor[HTML]{FFFFFF}126840} &
  \multicolumn{1}{r|}{\cellcolor[HTML]{FFFFFF}42280} &
  \multicolumn{1}{r|}{\cellcolor[HTML]{FFFFFF}0.4477} &
  \multicolumn{1}{r|}{\cellcolor[HTML]{FFFFFF}0.718939} &
   \\ \hline
\multicolumn{1}{|l|}{\cellcolor[HTML]{FFFFFF}Residuals} &
  \multicolumn{1}{r|}{\cellcolor[HTML]{FFFFFF}3365} &
  \multicolumn{1}{r|}{\cellcolor[HTML]{FFFFFF}317783541} &
  \multicolumn{1}{r|}{\cellcolor[HTML]{FFFFFF}94438} &
  \multicolumn{1}{l|}{\cellcolor[HTML]{FFFFFF}} &
  \multicolumn{1}{l|}{\cellcolor[HTML]{FFFFFF}} &
   \\ \hline
\multicolumn{7}{|l|}{\cellcolor[HTML]{FFFFFF}Signif. codes:  0 ‘***’ 0.001 ‘**’ 0.01 ‘*’ 0.05 ‘.’ 0.1 ‘ ’ 1} \\ \hline
\end{tabular}
\caption{ANOVA table for \textbf{H3} showing the influence of the independent variables on the dependent variable ``time'' together with their interaction effects.}
\end{table}


\begin{table}[]
\begin{tabular}{|
>{\columncolor[HTML]{FFFFFF}}l 
>{\columncolor[HTML]{FFFFFF}}r 
>{\columncolor[HTML]{FFFFFF}}r 
>{\columncolor[HTML]{FFFFFF}}r 
>{\columncolor[HTML]{FFFFFF}}r 
>{\columncolor[HTML]{FFFFFF}}r 
>{\columncolor[HTML]{FFFFFF}}r |}
\hline
\multicolumn{1}{|l|}{\cellcolor[HTML]{FFFFFF}\textbf{Response Time}} &
  \multicolumn{1}{l|}{\cellcolor[HTML]{FFFFFF}Df} &
  \multicolumn{1}{l|}{\cellcolor[HTML]{FFFFFF}Sum Sq} &
  \multicolumn{1}{l|}{\cellcolor[HTML]{FFFFFF}Mean Sq} &
  \multicolumn{1}{l|}{\cellcolor[HTML]{FFFFFF}F value} &
  \multicolumn{1}{l|}{\cellcolor[HTML]{FFFFFF}Pr(\textgreater{}F)} &
  \multicolumn{1}{l|}{\cellcolor[HTML]{FFFFFF}Significance} \\ \hline
\multicolumn{1}{|l|}{\cellcolor[HTML]{FFFFFF}network\_enc} &
  \multicolumn{1}{r|}{\cellcolor[HTML]{FFFFFF}1} &
  \multicolumn{1}{r|}{\cellcolor[HTML]{FFFFFF}1731835} &
  \multicolumn{1}{r|}{\cellcolor[HTML]{FFFFFF}1731835} &
  \multicolumn{1}{r|}{\cellcolor[HTML]{FFFFFF}30.8822} &
  \multicolumn{1}{r|}{\cellcolor[HTML]{FFFFFF}2.95E-08} &
  \cellcolor[HTML]{FFFFFF}*** \\ \hline
\multicolumn{1}{|l|}{\cellcolor[HTML]{FFFFFF}task\_type} &
  \multicolumn{1}{r|}{\cellcolor[HTML]{FFFFFF}1} &
  \multicolumn{1}{r|}{\cellcolor[HTML]{FFFFFF}8841140} &
  \multicolumn{1}{r|}{\cellcolor[HTML]{FFFFFF}8841140} &
  \multicolumn{1}{r|}{\cellcolor[HTML]{FFFFFF}157.656} &
  \multicolumn{1}{r|}{\cellcolor[HTML]{FFFFFF}\textless 2.2E-16} &
  \cellcolor[HTML]{FFFFFF}*** \\ \hline
\multicolumn{1}{|l|}{\cellcolor[HTML]{FFFFFF}group} &
  \multicolumn{1}{r|}{\cellcolor[HTML]{FFFFFF}1} &
  \multicolumn{1}{r|}{\cellcolor[HTML]{FFFFFF}1036281} &
  \multicolumn{1}{r|}{\cellcolor[HTML]{FFFFFF}1036281} &
  \multicolumn{1}{r|}{\cellcolor[HTML]{FFFFFF}18.4791} &
  \multicolumn{1}{r|}{\cellcolor[HTML]{FFFFFF}1.77E-05} &
  \cellcolor[HTML]{FFFFFF}*** \\ \hline
\multicolumn{1}{|l|}{\cellcolor[HTML]{FFFFFF}network\_enc:task\_type} &
  \multicolumn{1}{r|}{\cellcolor[HTML]{FFFFFF}1} &
  \multicolumn{1}{r|}{\cellcolor[HTML]{FFFFFF}2355561} &
  \multicolumn{1}{r|}{\cellcolor[HTML]{FFFFFF}2355561} &
  \multicolumn{1}{r|}{\cellcolor[HTML]{FFFFFF}42.0046} &
  \multicolumn{1}{r|}{\cellcolor[HTML]{FFFFFF}1.04E-10} &
  \cellcolor[HTML]{FFFFFF}*** \\ \hline
\multicolumn{1}{|l|}{\cellcolor[HTML]{FFFFFF}network\_enc:group} &
  \multicolumn{1}{r|}{\cellcolor[HTML]{FFFFFF}1} &
  \multicolumn{1}{r|}{\cellcolor[HTML]{FFFFFF}67782} &
  \multicolumn{1}{r|}{\cellcolor[HTML]{FFFFFF}67782} &
  \multicolumn{1}{r|}{\cellcolor[HTML]{FFFFFF}1.2087} &
  \multicolumn{1}{r|}{\cellcolor[HTML]{FFFFFF}2.72E-01} &
  \multicolumn{1}{l|}{\cellcolor[HTML]{FFFFFF}} \\ \hline
\multicolumn{1}{|l|}{\cellcolor[HTML]{FFFFFF}task\_type:group} &
  \multicolumn{1}{r|}{\cellcolor[HTML]{FFFFFF}1} &
  \multicolumn{1}{r|}{\cellcolor[HTML]{FFFFFF}342754} &
  \multicolumn{1}{r|}{\cellcolor[HTML]{FFFFFF}342754} &
  \multicolumn{1}{r|}{\cellcolor[HTML]{FFFFFF}6.112} &
  \multicolumn{1}{r|}{\cellcolor[HTML]{FFFFFF}0.01348} &
  * \\ \hline
\multicolumn{1}{|l|}{\cellcolor[HTML]{FFFFFF}network\_enc:task\_type:group} &
  \multicolumn{1}{r|}{\cellcolor[HTML]{FFFFFF}1} &
  \multicolumn{1}{r|}{\cellcolor[HTML]{FFFFFF}15681} &
  \multicolumn{1}{r|}{\cellcolor[HTML]{FFFFFF}15681} &
  \multicolumn{1}{r|}{\cellcolor[HTML]{FFFFFF}0.2796} &
  \multicolumn{1}{r|}{\cellcolor[HTML]{FFFFFF}0.59698} &
  \multicolumn{1}{l|}{\cellcolor[HTML]{FFFFFF}} \\ \hline
\multicolumn{1}{|l|}{\cellcolor[HTML]{FFFFFF}Residuals} &
  \multicolumn{1}{r|}{\cellcolor[HTML]{FFFFFF}3377} &
  \multicolumn{1}{r|}{\cellcolor[HTML]{FFFFFF}189377703} &
  \multicolumn{1}{r|}{\cellcolor[HTML]{FFFFFF}56079} &
  \multicolumn{1}{l|}{\cellcolor[HTML]{FFFFFF}} &
  \multicolumn{1}{l|}{\cellcolor[HTML]{FFFFFF}} &
  \multicolumn{1}{l|}{\cellcolor[HTML]{FFFFFF}} \\ \hline
\multicolumn{7}{|l|}{\cellcolor[HTML]{FFFFFF}Signif. codes:  0 ‘***’ 0.001 ‘**’ 0.01 ‘*’ 0.05 ‘.’ 0.1 ‘ ’ 1} \\ \hline
\end{tabular}
\caption{ANOVA table for \textbf{H4} showing the influence of the independent variables on the dependent variable ``time'' together with their interaction effects.}
\end{table}
\begin{table}[]
\begin{tabular}{|
>{\columncolor[HTML]{FFFFFF}}l 
>{\columncolor[HTML]{FFFFFF}}r 
>{\columncolor[HTML]{FFFFFF}}r 
>{\columncolor[HTML]{FFFFFF}}r 
>{\columncolor[HTML]{FFFFFF}}r 
>{\columncolor[HTML]{FFFFFF}}r 
>{\columncolor[HTML]{FFFFFF}}r |}
\hline
\multicolumn{1}{|l|}{\cellcolor[HTML]{FFFFFF}\textbf{Response Time}} &
  \multicolumn{1}{l|}{\cellcolor[HTML]{FFFFFF}Df} &
  \multicolumn{1}{l|}{\cellcolor[HTML]{FFFFFF}Sum Sq} &
  \multicolumn{1}{l|}{\cellcolor[HTML]{FFFFFF}Mean Sq} &
  \multicolumn{1}{l|}{\cellcolor[HTML]{FFFFFF}F value} &
  \multicolumn{1}{l|}{\cellcolor[HTML]{FFFFFF}Pr(\textgreater{}F)} &
  \multicolumn{1}{l|}{\cellcolor[HTML]{FFFFFF}Significance} \\ \hline
\multicolumn{1}{|l|}{\cellcolor[HTML]{FFFFFF}network\_enc} &
  \multicolumn{1}{r|}{\cellcolor[HTML]{FFFFFF}1} &
  \multicolumn{1}{r|}{\cellcolor[HTML]{FFFFFF}1357060} &
  \multicolumn{1}{r|}{\cellcolor[HTML]{FFFFFF}1357060} &
  \multicolumn{1}{r|}{\cellcolor[HTML]{FFFFFF}61.7501} &
  \multicolumn{1}{r|}{\cellcolor[HTML]{FFFFFF}5.25E-15} &
  \cellcolor[HTML]{FFFFFF}*** \\ \hline
\multicolumn{1}{|l|}{\cellcolor[HTML]{FFFFFF}task\_type} &
  \multicolumn{1}{r|}{\cellcolor[HTML]{FFFFFF}3} &
  \multicolumn{1}{r|}{\cellcolor[HTML]{FFFFFF}213354} &
  \multicolumn{1}{r|}{\cellcolor[HTML]{FFFFFF}71118} &
  \multicolumn{1}{r|}{\cellcolor[HTML]{FFFFFF}3.2361} &
  \multicolumn{1}{r|}{\cellcolor[HTML]{FFFFFF}0.02134} &
  \cellcolor[HTML]{FFFFFF}* \\ \hline
\multicolumn{1}{|l|}{\cellcolor[HTML]{FFFFFF}group} &
  \multicolumn{1}{r|}{\cellcolor[HTML]{FFFFFF}1} &
  \multicolumn{1}{r|}{\cellcolor[HTML]{FFFFFF}346104} &
  \multicolumn{1}{r|}{\cellcolor[HTML]{FFFFFF}346104} &
  \multicolumn{1}{r|}{\cellcolor[HTML]{FFFFFF}15.7487} &
  \multicolumn{1}{r|}{\cellcolor[HTML]{FFFFFF}7.39E-05} &
  \cellcolor[HTML]{FFFFFF}*** \\ \hline
\multicolumn{1}{|l|}{\cellcolor[HTML]{FFFFFF}network\_enc:task\_type} &
  \multicolumn{1}{r|}{\cellcolor[HTML]{FFFFFF}3} &
  \multicolumn{1}{r|}{\cellcolor[HTML]{FFFFFF}75322} &
  \multicolumn{1}{r|}{\cellcolor[HTML]{FFFFFF}25107} &
  \multicolumn{1}{r|}{\cellcolor[HTML]{FFFFFF}1.1425} &
  \multicolumn{1}{r|}{\cellcolor[HTML]{FFFFFF}3.30E-01} &
  \cellcolor[HTML]{FFFFFF} \\ \hline
\multicolumn{1}{|l|}{\cellcolor[HTML]{FFFFFF}network\_enc:group} &
  \multicolumn{1}{r|}{\cellcolor[HTML]{FFFFFF}1} &
  \multicolumn{1}{r|}{\cellcolor[HTML]{FFFFFF}2920} &
  \multicolumn{1}{r|}{\cellcolor[HTML]{FFFFFF}2920} &
  \multicolumn{1}{r|}{\cellcolor[HTML]{FFFFFF}0.1329} &
  \multicolumn{1}{r|}{\cellcolor[HTML]{FFFFFF}7.16E-01} &
  \multicolumn{1}{l|}{\cellcolor[HTML]{FFFFFF}} \\ \hline
\multicolumn{1}{|l|}{\cellcolor[HTML]{FFFFFF}task\_type:group} &
  \multicolumn{1}{r|}{\cellcolor[HTML]{FFFFFF}3} &
  \multicolumn{1}{r|}{\cellcolor[HTML]{FFFFFF}5082} &
  \multicolumn{1}{r|}{\cellcolor[HTML]{FFFFFF}1694} &
  \multicolumn{1}{r|}{\cellcolor[HTML]{FFFFFF}0.0771} &
  \multicolumn{1}{r|}{\cellcolor[HTML]{FFFFFF}0.97239} &
   \\ \hline
\multicolumn{1}{|l|}{\cellcolor[HTML]{FFFFFF}network\_enc:task\_type:group} &
  \multicolumn{1}{r|}{\cellcolor[HTML]{FFFFFF}3} &
  \multicolumn{1}{r|}{\cellcolor[HTML]{FFFFFF}36402} &
  \multicolumn{1}{r|}{\cellcolor[HTML]{FFFFFF}12134} &
  \multicolumn{1}{r|}{\cellcolor[HTML]{FFFFFF}0.5521} &
  \multicolumn{1}{r|}{\cellcolor[HTML]{FFFFFF}0.64671} &
  \multicolumn{1}{l|}{\cellcolor[HTML]{FFFFFF}} \\ \hline
\multicolumn{1}{|l|}{\cellcolor[HTML]{FFFFFF}Residuals} &
  \multicolumn{1}{r|}{\cellcolor[HTML]{FFFFFF}3262} &
  \multicolumn{1}{r|}{\cellcolor[HTML]{FFFFFF}71687780} &
  \multicolumn{1}{r|}{\cellcolor[HTML]{FFFFFF}21977} &
  \multicolumn{1}{l|}{\cellcolor[HTML]{FFFFFF}} &
  \multicolumn{1}{l|}{\cellcolor[HTML]{FFFFFF}} &
  \multicolumn{1}{l|}{\cellcolor[HTML]{FFFFFF}} \\ \hline
\multicolumn{7}{|l|}{\cellcolor[HTML]{FFFFFF}Signif. codes:  0 ‘***’ 0.001 ‘**’ 0.01 ‘*’ 0.05 ‘.’ 0.1 ‘ ’ 1} \\ \hline
\end{tabular}
\caption{ANOVA table for \textbf{H5} showing the influence of the independent variables on the dependent variable ``time'' together with their interaction effects.}
\end{table}

\end{document}